\documentclass{workshop}
\usepackage{epsfig}
\begin{document}
\begin{frontmatter}
\title{Neutral Pions and $\eta$ Mesons as Probes of the Expanding Hadronic
Fireball in Nucleus--Nucleus Collisions at SIS\thanksref{talk}}
\thanks[talk]{Presented at the IV. TAPS Workshop, Mont Sainte Odile, France,
September 1997}
\author{Ralf Averbeck}
\address{Gesellschaft f{\"u}r Schwerionenforschung, Planckstr. 1,
D-64291 Darmstadt, Germany}
\begin{abstract}
The production of $\pi^0$ and $\eta$ mesons in collisions of light, 
intermediate--mass, and heavy nuclei at beam energies between 0.8$A$~GeV
and 2.0$A$~GeV is discussed in the framework of a model assuming equilibrium
at hadrochemical and thermal freeze--out, respectively.
From the relative particle yields temperatures $T_C$ of 56~--~90~MeV and 
baryon chemical potentials $\mu_B$ of 780~--~675~MeV are deduced at chemical 
freeze--out.
Midrapidity spectra of $\pi^0$ and $\eta$ mesons are consistent with the
obtained temperatures if, in addition, the measured transverse expansion 
of the collision zone is taken into account.
In contrast to results from similar analyses of AGS and SPS data the
deduced freeze--out parameters are far below the phase boundary between
a hadron gas and a quark--gluon plasma.
\end{abstract}
\end{frontmatter}

\section{Introduction}

During the last two decades the investigation of the properties of hadronic
matter at high density and high temperature has become one of the key
issues of nuclear physics.
Relativistic nucleus--nucleus collisions offer the unique possibility to
study nuclear matter under such extreme conditions.
At incident energies around 1$A$~GeV the formation of a hot and dense 
reaction zone, the so--called nuclear fireball, has been verified 
experimentally \cite{STO86}.
According to model calculations, nuclear matter can be compressed for time
spans of $\approx$10~--~15~fm/c to densities of $\approx$2~--~3 times the
nuclear groundstate density \cite{CAS90,AIC91,MAR94,BAS95}.
Simultaneously, temperatures up to $\approx$100~MeV may be reached and a 
substantial fraction of the nucleons participating in the collision is 
excited to heavier, short--lived resonance states which decay predominantly 
by meson emission \cite{EHE93,MET93,MOS93}.
Hence, the fireball at SIS--energies represents a system of interacting 
nucleons, resonances, and mesons.
Under these conditions the onset of the partial restoration of chiral 
symmetry, which is a fundamental symmetry of quantum chromodynamics,
could manifest itself in modifications of hadron properties inside the 
fireball.
At beam energies of $\approx$10$A$~GeV, as they are available at the 
AGS accelerator facility at BNL (Brookhaven), even higher baryon densities 
and temperatures are reached in the initial stage of the fireball
\cite{KAH96}.
In addition to the chiral phase transition one may approach in those 
collisions already the phase transition from a hadron gas to a quark--gluon
plasma due to the deconfinement of quarks and gluons.
In such a scenario the fireball would no longer be purely hadronic during
the whole collision process but the initial stage should be a partonic
phase until hadronization takes place.
At the SPS accelerator facility at CERN (Geneva), where the highest energy 
nuclear collisions can be studied using 158$A$~GeV Pb beams, the possibility 
exists that the phase boundary to quark matter has been crossed already, 
although evidence for the phase transition is still rather tentative 
\cite{ABR96}.

It is the common goal of numerous experiments at SIS, AGS, and SPS to
investigate the hadronic or partonic fireball, respectively.
Since each relativistic nucleus--nucleus collision, irrespective of the
initial beam energy, develops a hadronic phase in its final stage, the
importance of studying the purely hadronic fireball as it is 
formed in the SIS--energy regime is evident.
Various observables, e.g. particle yields, spectra, and correlations
are sensitive to different parameters of the fireball.
Only the combination of all observables may finally give a complete picture 
of $A$~GeV heavy--ion collisions which up to now are not fully understood.

In this paper we concentrate on $\pi^0$ and $\eta$--meson production
as probes of the fireball.
Chapter~2 summarizes briefly some basic features of the time evolution
of relativistic nucleus--nucleus collisions as far as meson production
is concerned. 
On the basis of the presently available systematics of $\pi^0$ and
$\eta$--meson production the motivation for a thermal--model ansatz to
describe hadronic matter at freeze--out is given.
In the chapters~3 and~4 the model of a hadron gas in thermal and 
hadrochemical equilibrium is described in detail and applied to the
fireball as observed at SIS.
After fixing the model parameters by experimental data, the part of the
nuclear--matter phase diagram probed by hadronic observables at SIS is
determined and compared to corresponding results from AGS and SPS.
Furthermore, the chemical composition of the fireball at freeze--out is 
estimated and the consistency of particle yields and spectra is studied.

\section{$\pi^0$ and $\eta$ Mesons from the Fireball at SIS}

\subsection{Nucleus--Nucleus Collisions at SIS}

Numerous investigations of relativistic nucleus--nucleus reactions have
shown that for central collisions at SIS the relative motion of target 
and projectile matter in the reaction zone almost comes to rest.
While for heavy target--projectile combinations at moderate energies 
an isotropically expanding source at midrapidity has been observed 
\cite{REI97}, indications for incomplete stopping have been found
in rapidity--density distributions for energies above 
$\approx$ 1$A$~GeV \cite{HON98}.
The size of the fireball thus arising depends on the centrality of the
collision.
Following the participant--spectator model, the geometrical overlap of the 
two nuclei determines the number of nucleons $A_{part}$ which are directly 
involved.
In addition the energy available in the nucleon--nucleon system
$\sqrt{s} - 2 m_N$ can be obtained from the projectile energy, so that the
initial conditions of the system are fixed.
In the subsequent development the energy is transformed into the excitation
of thermal and collective degrees of freedom:
it turns into heat and provides the mass stored in the resonance states,
and it builds up compression and produces the flow of the expanding matter.

In a simplified picture the mesons are trapped during the compression phase
in a cyclic process of generation, absorption, and re--emission, exemplified
for nucleons $N$, $\pi$ mesons, and $\Delta$ resonances by 
$NN \rightleftharpoons N\Delta \rightleftharpoons NN\pi$.
Within this approach the bulk of the mesons are released only with the onset 
of the expansion phase, when due to the decreasing matter density mesons and 
baryons decouple.
This implies that hadronic probes reflect only the final state
of nucleus--nucleus collisions and do not provide direct insight into the 
early, high--density phase of the collision as leptonic probes would do
since they are not affected by the strong interaction. 
The moment in the expansion process when particles cease to interact 
strongly is called freeze--out.
One distinguishes between chemical freeze--out and thermal 
freeze--out which correspond to those points in time when the relative 
abundances or the momentum distributions of the particles, 
respectively, are no longer changed.
Obviously only inelastic collisions of nucleons, resonances, and mesons
can change the relative particle yields while the momentum distributions
are governed by the total interaction cross sections of these particles
in the fireball.
Therefore, chemical freeze--out does not occur after thermal freeze--out.
Even in this simplified sudden freeze--out scenario one has to make sure
that for the determination of freeze--out parameters only such observables
are combined with each other which are fixed approximately simultaneously.
In our ansatz particle yields reflect chemical freeze--out
while momentum distributions reflect thermal freeze--out.
The latter, however, are not purely thermal spectra in the sense that they 
follow Boltzmann distributions but they are modified by resonance decays
and reflect in addition collective flow phenomena.

It is an interesting question to what extent the fireball at freeze--out 
can be described in terms of thermal and hadrochemical equilibrium.
Equilibrium could be achieved through frequent scattering of particles in 
the fireball and it would be reached without any doubt if only the lifetime
of the fireball would be sufficiently long.
It has been shown that at AGS and SPS the hadronic observables including
strange particles are in quite good agreement with such a scenario 
\cite{CLE93,BRA95,BRA96,SOL97}, while at SIS the situation is still unclear.
Only recently it has been reported that midrapidity momentum spectra of
nonstrange particles measured in central $^{58}$Ni~+~$^{58}$Ni collisions are 
consistent with thermal equilibrium in the energy regime from 1$A$~GeV 
to 2$A$~GeV if in addition collective radial flow is taken into account
\cite{HON98}.
For the same reactions also agreement with chemical equilibrium has been
claimed \cite{HON96}.
This analysis, however, was based on proton, deuteron, thermal charged pion,
and $\Delta(1232)$--resonance yields.
The latter were obtained by decomposing the pion momentum spectra into
a thermal component and into a component from resonance decay.
In this paper we follow a different strategy:
in order to keep the possibility to distinguish chemical and thermal 
freeze--out as individual stages in the time evolution of relativistic
nucleus--nucleus collisions we strictly confine ourselves to an isolated 
discussion of relative particle yields and momentum distributions, 
respectively, without combining these informations.
In order to keep the thermal model as simple as possible we do not take 
into account particles with open strangeness since at SIS energies their 
production rate is very small.
Furthermore, an additional model parameter, the strangeness chemical 
potential $\mu_S$, which, however, could be fixed by requesting 
strangeness conservation, would be necessary to incorporate strangeness
consistently into the thermal model.
With respect to strange particles, only recently Cleymans and co--workers
have claimed that at 1.93$A$~GeV the particle yields including the strange
hadrons agree with a complete chemical equilibrium scenario at freeze--out
\cite{CLE97}.
This would mean that not only the nonstrange and strange hadrons among 
themselves could be equilibrated seperately, but even equilibrium between 
both the strange and the nonstrange sectors would have been established.

\subsection{Results from $\pi^0$ and $\eta$ meson production experiments}

\begin{figure}[tbh]
\begin{center}
\mbox{\epsfig{file=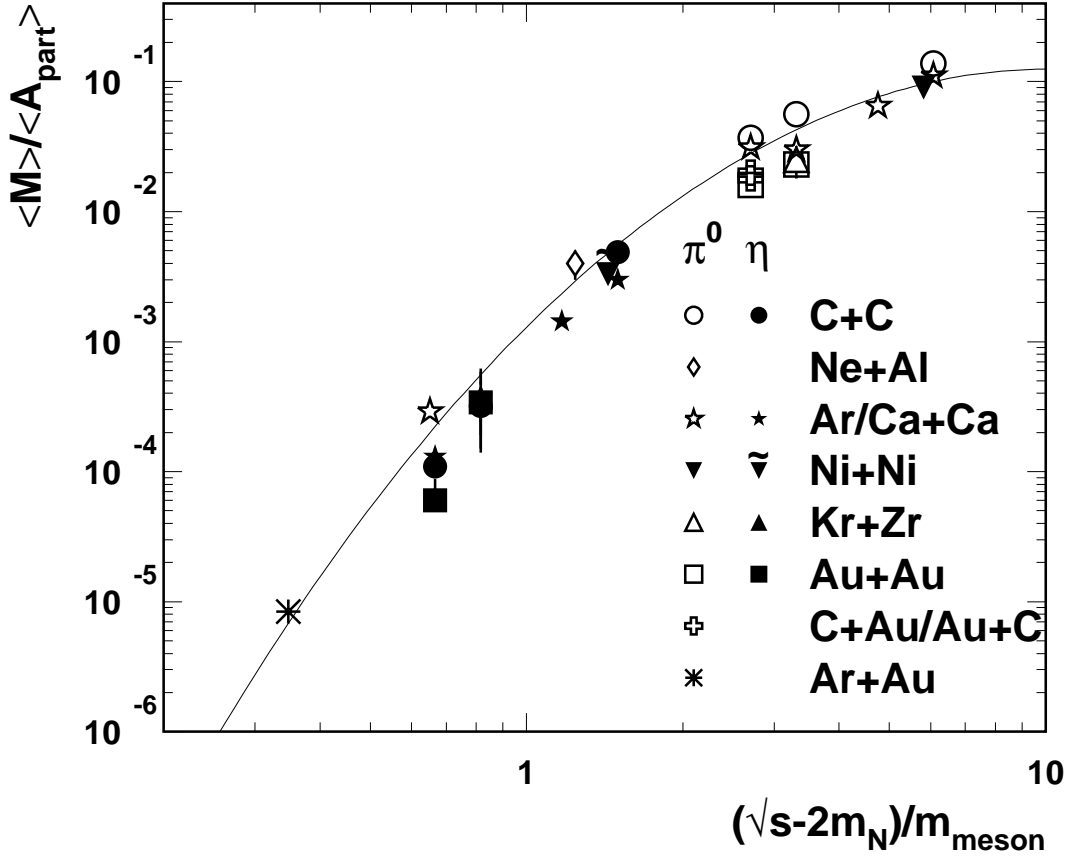,width=\textwidth}}
\end{center}
\caption[]{Average $\pi^0$ and $\eta$--meson multiplicities per average
number of participants in nucleus--nucleus collisions as a function of
the energy available in the nucleon--nucleon system. The data are taken from
\cite{PFE93,BER94,SCH94,MET94,SCH95,APP97,AVE97,MAR97,MAR98,VOG98,WOL98}. The 
line represents a fit, taking into account also charged--meson production in 
addition \cite{MET93}.}
\label{metag}
\end{figure}

The neutral $\pi^0$ and $\eta$ mesons which have large decay branches 
into pairs of photons represent unique probes for the expanding fireball 
and therefore provide the basis for our discussion.
Pions and $\eta$ mesons are the only mesons with substantial production 
rates at SIS energies.
While the pionic degree of freedom is also covered by the spectroscopy of 
the charged members of the multiplet, it is only through gamma-ray 
spectroscopy that the $\eta$ meson becomes observable in nucleus--nucleus 
collisions.
These two neutral mesons can be simultaneously identified by a two-photon 
invariant-mass analysis of coincident photon pairs. 
During the last almost ten years the TAPS collaboration performed a series 
of systematic meson--production experiments covering incident energies from 
0.1$A$~GeV to 2.0$A$~GeV. 
These measurements have established an extended data base for $\pi^0$ and 
$\eta$ production in light ($^{12}$C + $^{12}$C), intermediate ($^{40}$Ar, 
$^{40}$Ca + $^{nat}$Ca, $^{58}$Ni + $^{58}$Ni, $^{86}$Kr + $^{nat}$Zr) and 
heavy systems ($^{197}$Au + $^{197}$Au) 
\cite{PFE93,BER94,SCH94,MET94,SCH95,APP97,AVE97,MAR97,MAR98,VOG98,WOL98}.
The primary information are the spectral distributions of the mesons,
measured down to zero transverse momentum in narrow bins around midrapidity, 
and the meson yields, which are determined in the same rapidity intervals
and are then extrapolated to the full solid angle.
All data concerning the experiments $^{40}$Ar + $^{nat}$Ca at 0.18$A$~GeV
\cite{MAR98}, $^{58}$Ni + $^{58}$Ni at 1.9$A$~GeV \cite{APP97}, and 
$^{40}$Ca + $^{nat}$Ca at 2.0$A$~GeV \cite{VOG98} presented in this paper 
are not yet submitted for publication and should therefore be regarded as 
preliminary results.

The first important result is obtained from the systematic investigation
of the meson yields.
Fig.~\ref{metag} shows the average meson multiplicities $<M>$, measured 
in inclusive reactions of various systems, normalized to the average
number of participating nucleons $<A_{part}>$ as a function of the 
energy $\sqrt{s}-2m_N$ available in the $NN$ system normalized to the 
corresponding meson mass.
A steep rise of the relative yields with increasing energy is visible.
In addition, a weaker dependence on the size of the collision system is 
observed with the clear tendency for smaller inclusive yields in the 
heavier systems. 
This has also been observed in charged--pion production experiments
\cite{MUE97,PEL97}.
In first approximation all data points fall onto a smooth curve 
indicating that the meson-production probabilities mainly depend 
on the available energy. 
This is quite remarkable because $\pi^0$ and $\eta$ production proceed 
through different baryon resonances: $\pi^0$ mesons mainly originate from 
$\Delta(1232)$--resonance decays, while the heavier $\eta$ mesons essentially 
come from the N(1535) resonance which, at SIS energies, is the only
significantly populated baryon resonance with a large decay
channel into $\eta$ mesons.
The fact that the specific production mechanism seems not to influence
the observed meson yields significantly may be interpreted as a first
indication for meson emission from an equilibrated source.

\begin{figure}[tbh]
\begin{center}
\mbox{\epsfig{file=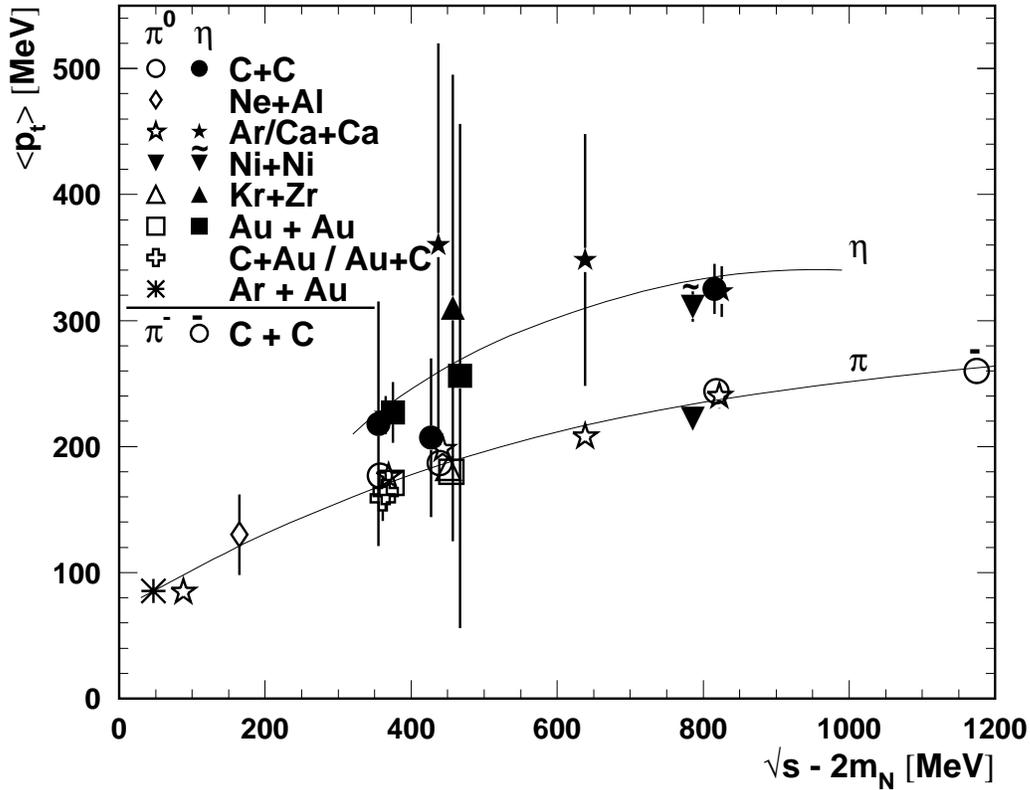,width=\textwidth}}
\end{center}
\caption[]{Average $\pi^0$ and $\eta$--meson transverse momenta measured
in narrow rapidity intervals around midrapidity as a function of the energy
available in the nucleon--nucleon system. The data are taken from the same
experiments as listed in Fig.~\ref{metag}. The data points around available
energies of 365~MeV (0.8$A$~GeV beam energy) and 447~MeV (1.0$A$~GeV beam
energy) are shifted slightly on the energy axis with respect to their
nominal position to resolve the error bars. Lines interpolating the $\pi^0$
and $\eta$ data, respectively, are drawn to guide the eyes. A $\pi^-$ point
taken from \cite{SIM86} has been added to the data set.}
\label{ptsys}
\end{figure}

The second important result concerns the average transverse momenta $<p_t>$.
Fig.~\ref{ptsys} shows $<p_t>$ of $\pi^0$ and $\eta$ mesons as a function
of the energy available in the $NN$ system.
Although the experimental uncertainties for the $\eta$ data are large 
compared to the $\pi^0$ data, the $\eta$ transverse-momentum spectra 
reveal at a given energy a tendency towards larger mean values $<p_t>$
which is manifest around 2.0$A$~GeV beam energy.
To give a measure for the expected order of magnitude of $<p_t>$, it
can be noted that for the decay of $\Delta(1232)$ and $N(1535)$ 
resonances at rest one would get 
$<p_t>_{\pi^0} = 225$~MeV\footnote{Throughout this paper we follow the usual 
convention $c = \hbar = 1$.} and $<p_t>_{\eta} = 182$~MeV, respectively.
For both mesons $<p_t>$ increases with the available energy which could
be interpreted in terms of a considerable heating of the collision zone
or a gradually growing population of the full resonance--mass 
distributions.
For $\pi^0$ mesons the rise of the average transverse momenta becomes 
smaller with increasing energy which is supported by a measurement of 
$<p_t>$ for $\pi^-$ mesons produced at 3.37$A$~GeV bombarding energy
\cite{SIM86}.
The steep rise of the meson--production yields on the one hand and the 
onset of a saturation of average transverse momenta on the other hand 
clearly indicate that with increasing bombarding energy the fraction of 
available energy converted into massive particles becomes larger compared 
to the chaotic motion of particles in the fireball.

\begin{figure}[tbh]
\begin{center}
\mbox{\epsfig{file=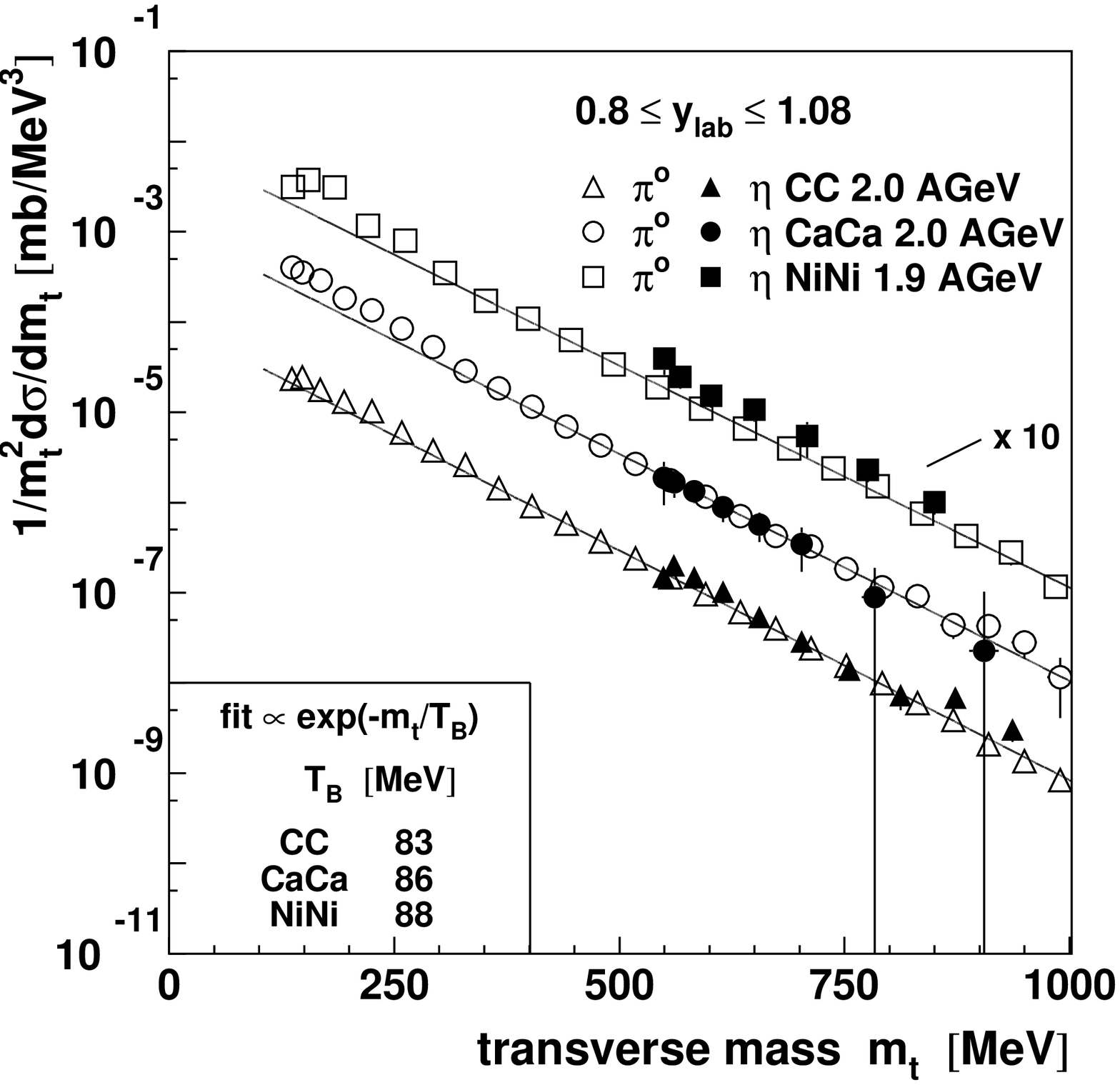,width=\textwidth}}
\end{center}
\caption[]{Transverse--mass spectra of $\pi^0$ and $\eta$--mesons as observed
in the systems $^{12}$C+$^{12}$C \cite{AVE97} and $^{40}$Ca+$^{nat}$Ca 
\cite{VOG98} at 2.0$A$~GeV beam energy and in $^{58}$Ni+$^{58}$Ni 
\cite{APP97} at 1.9$A$~GeV. The distributions are divided by the square
of the transverse mass. In this representation midrapidity particle spectra
from a source in thermal equilibrium are expected to exhibit a purely 
exponential spectrum. The solid lines represent exponential fits to the
$\pi^0$ data for $m_t \ge 400$~MeV.}
\label{mtscal}
\end{figure}

Finally, the distributions of the transverse mass $m_t$ of $\pi^0$ and
$\eta$ mesons are discussed.
At midrapidity the transverse mass of a particle is equivalent to its
total energy in the center--of--mass system.
Following \cite{STA92} the transverse--mass distribution of particles 
with mass $m$, emitted isotropically from a thermal source and 
characterized by a Boltzmann temperature $T_B$, can approximately be 
described by

\begin{equation}
\frac{1}{m_t^2}\,\frac{d\sigma}{dm_t} \propto 
\exp\left(\frac{-m_t}{T_B}\right) 
\;\;\mbox{with}\;\; m_t = \sqrt{m^2 + p_t^2}
\label{mtdis}
\end{equation}

In Fig.~\ref{mtscal} transverse--mass distributions of $\pi^0$ and $\eta$
mesons are plotted together with fits according to the parametrization 
given in Eq.~\ref{mtdis} for three different collision systems at incident 
energies near 2$A$~GeV.
Within each reaction the $\pi^0$ and $\eta$ spectra exhibit almost identical
inverse--slope parameters $T_B$ which in addition do not change significantly
with the mass of the colliding nuclei.
Such a behaviour, the so--called $m_t$ scaling, is well known in
high--energy physics \cite{BOU76,AGU91}.
Having in mind the different production mechanisms for both meson species
it is furthermore quite remarkable that for $m_t \ge m_\eta$ the $\pi^0$
and $\eta$ data roughly fall on one commen line for all systems.
This indicates that the energy required to produce a given transverse mass
completely determines the relative abundance of the meson species near 
midrapidity.
For low $m_t$, however, individual differences become apparent.
Fig.~\ref{mtscal} shows a systematic enhancement over the exponential rise 
extrapolated from the high--$m_t$ region if one goes from the light to
the heavier systems.
Possible mechanisms that have been suggested to account for this fact
involve enhanced pion rescattering through resonance states in the
heavier systems \cite{BAS93} and multiple pion decay of heavy 
resonances \cite{TEI97}.
The same observation, namely the $m_t$ scaling on an absolute scale
and the low--$m_t$ enhancement in heavier systems, is not only manifest
at 2$A$~GeV but has been recognized at various energies down to 
0.8$A$~GeV and seems to be a general feature of heavy--ion collisions
in the SIS--energy regime \cite{BER94,APP97,AVE97,MAR97,VOG98,WOL98}.
Although $m_t$ scaling does not prove that the meson source actually
has reached thermal or hadrochemical equilibrium, the particle spectra
of such an equilibrated source would be dominated by phase space as it
is observed.

In summary, the systematic study of $\pi^0$ and $\eta$ meson yields 
and momentum distributions reveals strong indications for the hypothesis
that these mesons may be emitted from a source in hadrochemical and
thermal equilibrium.

\section{Particle Yields and Chemical Freeze--Out in a Hadron--Gas Model} 

\subsection{Formalism of the Hadron--Gas Model}

In our ansatz we assume that at the point of chemical freeze--out the
fireball can be described in terms of an equilibrated hadron gas.
As constituents of this ideal gas we take into account nucleons, deuterons, 
pions, $\eta$ mesons, and all nonstrange baryon resonances up to a mass of 
1.76~GeV, which corresponds to $\sqrt{s}-m_N$ for the nucleon--nucleon system 
at 2$A$~GeV beam energy.

Nuclear matter in hadrochemical equilibrium has a particle composition 
characterized by a temperature $T_C$ and a baryon chemical potential $\mu_B$.
Considering the grand--canonical ensemble description of noninteracting
fermions and bosons the particle--number densities $\rho_i$ are given by
integrals over the particle momentum $p$:

\begin{equation} 
\rho_{i} = \frac{g_i}{2\pi^2}\;
           \int_0^\infty\frac{p^2\,dp}
           {exp\left[\left(E_{i}-\mu_{B} B_{i} \right)/T_C \right] \pm 1}
\label{grandcan}
\end{equation} 

where $g_i$ is the spin--isospin degeneracy, $E_i$ the total energy in the
local restframe, and $B_i$ the baryon number of the particle species $i$.
The $\pm$~sign refers to the different statistics for fermions or bosons,
respectively.

Eq.~\ref{grandcan} can not be applied to the fireball at SIS directly but
only after the inclusion of two modifications.
To account for the fact that the fireball occupies only a finite volume,
a surface correction has to be included.
For an estimate of this correction we assume a spherical volume with
radius $R$ giving a momentum dependent correction factor $f(pR)$ 
\cite{BAL70,JAQ84} which has to be multiplied with the integrand in 
Eq.~\ref{grandcan}:

\begin{equation}
f(pR) = 1 - \frac{3\pi}{4pR} + \frac{1}{\left(pR\right)^2}
\label{surface}
\end{equation} 

Introducing this correction leads to a decrease of individual
particle--number densities down to 60~--~70~\% of the uncorrected 
values for reasonable radius parameters.
Particle yield ratios, however, are much less affected by the correction
since there the correction almost cancels out. 
The radius parameter should be chosen such that the number of baryons
inside the spherical volume is in reasonable agreement with the average
number of nucleons participating in the collision.
We use a constant radius parameter of 5.5~fm for all beam energies.
If one assumes that chemical freeze--out may occur at baryon densities
in the range between 0.2 and 0.6~$\rho_0$ (see below) the corresponding 
freeze--out volume would then contain $\approx$23~-~70 baryons.
We verified that in our model analysis a variation of $R$ by $\pm$2~fm
changes the resulting temperatures and chemical potentials by less than 5~\%.

The second modification is related to the fact that resonances do not have 
a fixed mass but exhibit a broad mass distribution.
In contrast to thermal--model analyses of AGS and SPS data, a reasonable
treatment of mass distributions can not be neglected in the 1$A$~GeV
energy regime since in this case the energy available in the $NN$ system 
is small compared to the difference in mass between the nucleon 
and the resonances except for the $\Delta(1232)$ resonance.
That is the reason why the $\Delta(1232)$ is by far the most abundantly 
populated resonance and why all the other resonances are populated 
predominantly in the tails of their mass distributions towards lower 
masses.
Eq.~\ref{grandcan} therefore has to be folded with normalized mass 
distributions $A_i(m)$ which are $\delta$ functions for all particle
species except resonances.
The following scheme to parametrize resonance mass distributions is to 
a large extent adopted from \cite{TEI97}.

We distribute the resonance masses according to Lorentzian functions 
$A_R(m)$ which are determined by the mean resonance masses $m_R$ and 
the mass dependent total and partial decay widths $\Gamma(m)$:

\begin{equation}
A_R(m) = \frac{m^2\Gamma_{tot}(m)}
              {\left(m^2-m_R^2\right)^2 + m^2\Gamma_{tot}^2(m)}
\;\;\mbox{with}\;\;
\Gamma_{tot}=\Gamma_{1\pi}+\Gamma_{2\pi}+\Gamma_{\eta} 
\label{lorentz}
\end{equation}

We use the following parametrizations for the partial decay widths 
$\Gamma(m)$ where the mean masses $m_R$ and widths $\Gamma_R$ at 
$m_R$ are taken from \cite{BAR96}:
\begin{itemize}
\item For the pion decay width of the $\Delta(1232)$ resonance we
      adopt the parametrization given by Koch et al. \cite{KOC84}
      \begin{equation}
      \Gamma(q) = \Gamma_R \frac{m_\Delta}{m} \left(\frac{q}{q_R}\right)^3
      \left(\frac{q_R^2+\delta^2}{q^2+\delta^2}\right)^2
      \end{equation}
      where $m$ is the actual mass of the $\Delta(1232)$ resonance.
      $q$ and $q_R$ are the pion three momenta in the restframe of the 
      resonance with mass $m$ and $m_\Delta$, respectively.
      The parameter $\delta$ in the cutoff function has the value 
      $\delta = 300$~MeV.
\item The 1$\pi$ and $\eta$ decay widths of the higher baryon resonances are
      given by
      \begin{equation}
      \Gamma(q) = \Gamma_R B_{1\pi,\eta} \left(\frac{q}{q_R}\right)^{2l+1}
      \left(\frac{q_R^2+\delta^2}{q^2+\delta^2}\right)^{l+1}
      \end{equation}
      where $B_{1\pi,\eta}$ is the branching ratio of the resonance into
      the $1\pi$ or $\eta$ decay channel at $m_R$, respectively, $l$ is the 
      angular momentum of the emitted meson, and $q$ and $q_R$ are the momenta
      of the meson in the restframe of the decaying resonance as defined 
      above. In this case we use 
      \begin{equation}
      \delta^2 = \left(m_R-M_N-m_{\pi,\eta}\right)^2 + \frac{\Gamma_R^2}{4}
      \end{equation}
      It has to be noted that the only resonance decaying into
      the $\eta$ channel in our ansatz is the $N(1535)$ resonance.
\item The 2$\pi$ decay widths of higher baryon resonances are described
      in terms of one--step processes. This is in contrast to \cite{TEI97}
      where in a first step a higher baryon resonance decays into a
      $\Delta(1232)$ or $N(1440)$ resonance and a pion or into a nucleon
      and a $\rho$ or $\sigma$ meson, and where in a second step the
      intermediate resonance decays after propagation through the nuclear
      medium into a nucleon and a pion or into a nucleon and two pions.
      Since the population of higher resonances is small compared to the
      $\Delta(1232)$ population we simplify this procedure by treating the
      2$\pi$ decay as a direct process where the resonance decays into
      a nucleon and an object with angular momentum $l=0$ and twice the
      pion mass which subsequently decays into two pions. The partial
      decay widths can then be written in complete analogy to the
      1$\pi$ and $\eta$ decay width. 
\item Additional decay channels are neglected.
\end{itemize}

Furthermore, one could consider an excluded--volume correction to take into
account the hadron--hadron hard--core repulsion or in other words the
difference between a real and an ideal hadron gas.
Since such a correction does not play a significant role for particle--yield
ratios as demonstrated in \cite{SOL97} we stick to the ideal hadron--gas
model.
Another argument to omit the excluded volume correction is the fact 
that chemical freeze--out seems to occur at baryon densities of the order
of half the nuclear groundstate density or even less as discussed below.
The hadron gas therefore is rather dilute which justifies the assumption
of non--interacting particles at chemical freeze--out.

Finally, the isospin asymmetry in collisions of heavy nuclei, which has
to be considered if the model analysis for example relies on charged--pion 
yields, is not taken into account in the present analysis.
This is adequate since our approach is based on the isospin--independent
$\eta$--meson yields and the neutral--pion yields which in the isobar
model correspond to one third of the total pion yields irrespective of
the actual isospin asymmetry in the system.

In summary, the equation we finally use to evaluate the particle--number
densities within our hadron--gas model reads as follows:

\begin{equation} 
\rho_{i} = \frac{g_i}{2\pi^2}\;
           \int_0^\infty f(pR)p^2\,dp
           \int_0^\infty \frac{A_i(m)\,dm}
           {exp\left[\left(\sqrt{m^2+p^2}-\mu_{B} 
           B_{i} \right)/T_C \right] \pm 1}
\label{partdens}
\end{equation} 

Because we have parametrized the resonance--mass distributions using
mass--dependent partial decay widths we can easily calculate the 
particle--number densities of mesons originating from any individual
resonance decay, which turns out to be essential for the determination
of the model parameters from experimental data.
If $\rho_{1\pi,2\pi,\eta}^{res}$ is defined as the particle--number density 
of pions or $\eta$--mesons stemming from a specific decay channel of an 
individual resonance one obtains

\begin{equation} 
\rho_{1\pi,2\pi,\eta}^{res} = \frac{w g_i}{2\pi^2}\;
           \int_0^\infty f(pR)p^2\,dp
           \int_0^\infty
           \frac{\Gamma_{1\pi,2\pi,\eta}(m) A_i(m)\,dm}
           {\Gamma_{tot}(m) exp\left[\left(\sqrt{m^2+p^2}-\mu_{B} 
           B_{i} \right)/T_C \right] \pm 1}
\label{decayyield}
\end{equation} 

with $w=1$ in the case of one--pion or $\eta$--meson decay and $w=2$
for the two--pion decay of a resonance.

\subsection{Determination of the Model Parameters from Experimental Data}

\begin{table}
\caption[]{Mean inclusive $\pi^0$ multiplicities $<M_{\pi^0}>$ relative to
the average number of participating nucleons $<A_{part}>$ and mean inclusive
$\eta$--meson multiplicities $<M_{\eta}>$ relative to $<M_{\pi^0}>$ 
measured for various target--projectile combinations in the beam--energy range
from 0.8$A$~GeV to 2.0$A$~GeV. The system--size averaged values are given in 
addition. The 2.0$A$~GeV Ca and 1.9$A$~GeV Ni data are still preliminary. 
$^\ast$For the average the latter have been upscaled by $\approx$10~\% to 
2.0$A$~GeV using the fit of the energy dependence as shown in 
Fig.~\ref{metag}.}
\label{data}
\begin{center}
\begin{tabular}{|c|c|c|c|c|}
\hline
$E$ [$A$~GeV] & System & $\frac{<M_{\pi^0}>}{<A_{part}>}$ [\%] & 
$\frac{<M_{\eta}>}{<M_{\pi^0}>}$ [\%] & Reference \\
\hline
0.8 & $^{12}$C+$^{12}$C & 3.7 $\pm$ 0.3 & 0.31 $\pm$ 0.11 & \cite{AVE97} \\
0.8 & $^{40}$Ar+$^{nat}$Ca & 3.1 $\pm$ 0.5 & 0.41 $\pm$ 0.04 & \cite{MAR97} \\
0.8 & $^{197}$Au+$^{197}$Au & 1.6 $\pm$ 0.3 & 0.38 $\pm$ 0.10 & \cite{WOL98} \\
0.8 & $^{12}$C + $^{197}$Au & 1.8 $\pm$ 0.2 & --- & \cite{AVE97} \\
0.8 & $^{197}$Au + $^{12}$C & 2.0 $\pm$ 0.2 & --- & \cite{AVE97} \\
\hline
0.8 & Average & 2.6 $\pm$ 0.3 & 0.39 $\pm$ 0.06 &  \\
\hline
1.0 & $^{12}$C+$^{12}$C & 5.6 $\pm$ 0.4 & 0.57 $\pm$ 0.14 & \cite{AVE97} \\
1.0 & $^{40}$Ar+$^{nat}$Ca & 3.0 $\pm$ 0.3 & 1.3 $\pm$ 0.8 & 
\cite{BER94,SCH94} \\
1.0 & $^{86}$Kr+$^{nat}$Zr & 2.5 $\pm$ 0.7 & 1.3 $\pm$ 0.5 & 
\cite{BER94,SCH94,MET94} \\
1.0 & $^{197}$Au+$^{197}$Au & 2.3 $\pm$ 0.5 & 1.4 $\pm$ 0.5 & 
\cite{BER94,SCH94,MET94} \\
\hline
1.0 & Average & 4.0 $\pm$ 0.4 & 1.06 $\pm$ 0.41 &  \\
\hline
1.5 & $^{40}$Ar+$^{nat}$Ca & 6.5 $\pm$ 0.5 & 2.2 $\pm$ 0.4 & 
\cite{BER94,SCH94} \\
\hline
2.0 & $^{12}$C+$^{12}$C & 13.8 $\pm$ 1.4 & 3.6 $\pm$ 0.4 & \cite{AVE97} \\
2.0 & $^{40}$Ca+$^{nat}$Ca & 11.1 $\pm$ 1.1 & 2.7 $\pm$ 0.3 & \cite{VOG98} \\
1.9 & $^{58}$Ni+$^{58}$Ni & 8.6 $\pm$ 0.9 & 3.3 $\pm$ 0.4 & \cite{APP97} \\
\hline
2.0 & Average$^\ast$  & 11.3 $\pm$ 1.1 & 3.34 $\pm$ 0.37 & \\
\hline
\end{tabular}
\end{center}
\end{table}

For a given beam energy the values of the two parameters of the 
hadron--gas model -- the temperature $T_C$ and the baryon chemical
potential $\mu_B$ -- can be derived from at least two relative abundances 
of the constituents.
The $\pi^0$ and $\eta$--meson yields are particularly suited for such an 
analysis, as these particles unambiguously arise from the fireball.
We base our analysis on the system--size averaged ratios 
$<M_{\pi^0}>/<A_{part}>$ and $<M_{\eta}>/<M_{\pi^0}>$ as compiled in 
Tab.~\ref{data} (see also Fig.~\ref{metag}).
The measured ratios are related to the particle--number densities calculated
within the model by two equations:

\begin{eqnarray}
\left(\frac{<M_{\pi^0}>}{<A_{part}>}\right)_{exp} & = & \frac{\left(\rho_\pi + 
\sum\limits_{Res}\rho_{1\pi}^{Res} + \rho_{2\pi}^{Res}\right)/3}
{\rho_N + 2 \rho_D + \sum\limits_{Res}\rho_{Res}} \; \left(\mu_B,T_C\right) \\
& & \nonumber\\
\left(\frac{<M_{\eta}>}{<M_{\pi^0}>}\right)_{exp} & = & \frac{\rho_\eta + 
\rho_\eta^{N(1535)}}{\left(\rho_\pi + \sum\limits_{Res}\rho_{1\pi}^{Res} + 
\rho_{2\pi}^{Res}\right)/3} \; \left(\mu_B,T_C\right)
\end{eqnarray}

\begin{figure}[tbh]
\begin{center}
\mbox{\epsfig{file=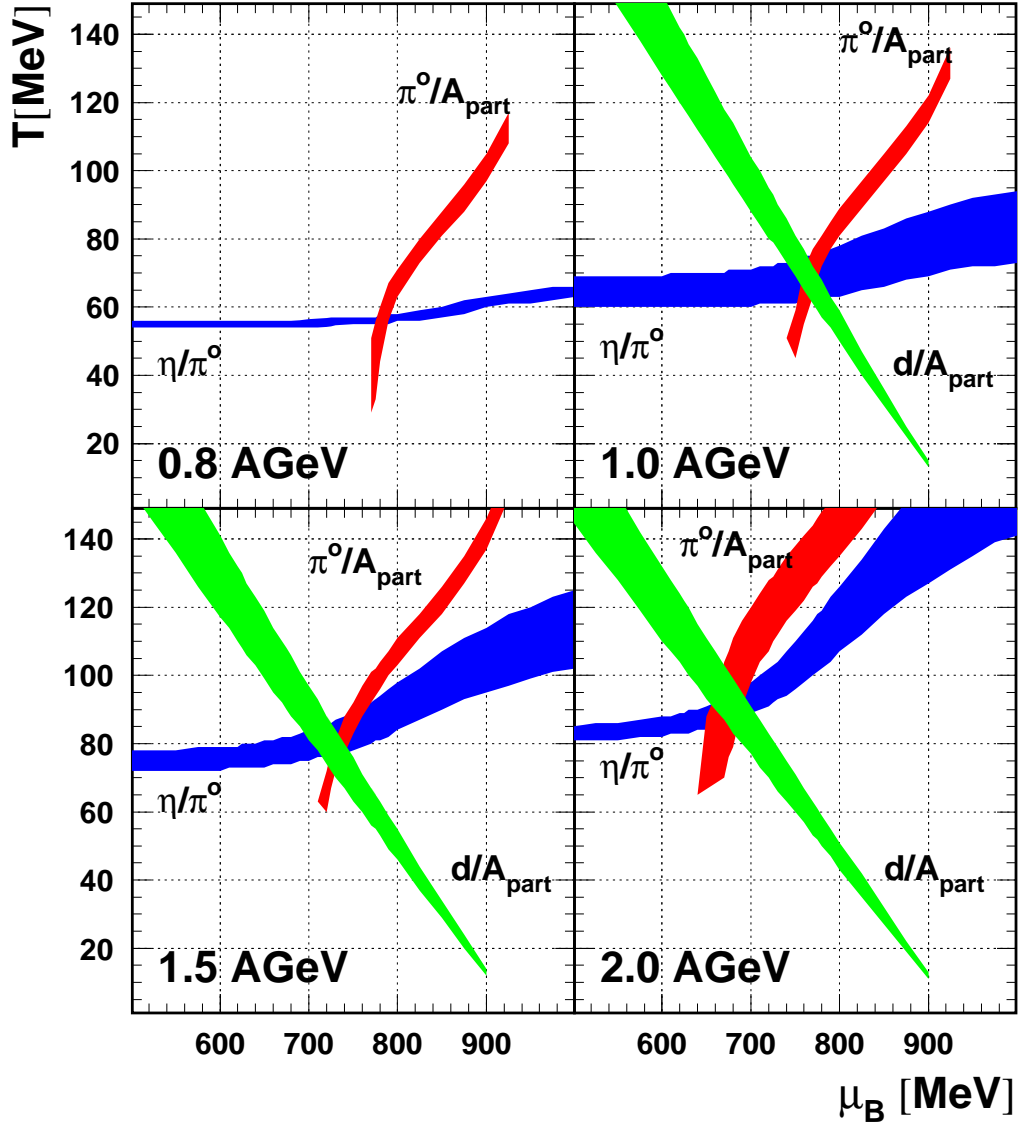,width=\textwidth}}
\end{center}
\caption[]{Determination of chemical freeze--out parameters: for a given
beam energy the two measured particle ratios $<M_{\pi^0}>/<A_{part}>$ and
$<M_{\eta}>/<M_{\pi^0}>$ define two bands in the parameter plane $(T_C,\mu_B)$ 
as described in the text. The four panels show these bands for beam energies
of 0.8, 1.0, 1.5, and 2.0$A$~GeV. The freeze--out parameters are determined 
by the $T_C$ and $\mu_B$ values where the two bands cross each other.
Except for the 0.8$A$~GeV case additional bands are shown which are defined 
by deuteron--to--nucleon ratios measured in central $^{58}$Ni+$^{58}$Ni 
collisions near the considered beam energies \cite{HON98}.}
\label{mut}
\end{figure}

To deduce $\mu_B$ and $T_C$ we determine within our model the chemical
potentials and temperatures for which the calculated ratios are in
agreement with the experimental values for $<M_{\pi^0}>/<A_{part}>$
and $<M_{\eta}>/<M_{\pi^0}>$, respectively.
Each ratio therefore defines an area in the parameter plane $T_C$
versus $\mu_B$.
Fig.~\ref{mut} shows the resulting areas for the four considered beam 
energies 0.8, 1.0, 1.5, and 2.0$A$~GeV..
The ratio $<M_{\pi^0}>/<A_{part}>$ is strongly correlated with $\mu_B$.
To increase $<M_{\pi^0}>/<A_{part}>$ smaller values for the chemical 
potential of the hadron gas are needed.
In contrast to that, the ratio $<M_{\eta}>/<M_{\pi^0}>$ is mainly correlated
with $T_C$.
With increasing $<M_{\eta}>/<M_{\pi^0}>$ larger values for the hadron--gas
temperature are necessary to obtain agreement between experimental and
calculated ratios.
For all beam energies the areas defined by the experimental boundary
conditions cross each other exactly once and therefore determine the
temperature and the baryon chemical potential of the hadron gas at
chemical freeze--out in an unambiguous way.
In addition, bands corresponding to deuteron--to--nucleon ratios 
measured by the FOPI collaboration \cite{HON98} in central 
$^{58}$Ni~+~$^{58}$Ni collisions at beam energies of 1.06$A$~GeV, 
1.45$A$~GeV, and 1.93$A$~GeV, respectively, are shown in Fig.~\ref{mut}.
Obviously also these ratios are in nice agreement with the chemical
freeze--out parameters obtained from inclusive $\pi^0$ and $\eta$--meson
yields.
The results are summarized in Tab.~\ref{resmut}.
The analysis reveals a systematic increase of the freeze--out temperature
from 56 to 90~MeV between 0.8 and 2.0$A$~GeV which is accompanied by a
reduction of the baryon chemical potential from 780 to 675~MeV. 
The corresponding baryon density at chemical freeze--out is
approximately 0.2~--~0.6 times the nuclear groundstate density which
is in reasonable agreement with results from other analyses even at 
higher energies \cite{BRA95,BRA96,HON96}.

\begin{table}
\caption{Chemical freeze--out parameters $T_C$ and $\mu_B$ obtained from
inclusive $\pi^0$ and $\eta$--meson yields for the four considered beam 
energies.}
\label{resmut}
\begin{center}
\begin{tabular}{|c|c|c|}
\hline
$E$ [$A$~GeV] & $T_C$ [MeV] & $\mu_B$ [MeV] \\
\hline
0.8 & 56 $\pm$ 2 & 780 $\pm$ 10 \\
1.0 & 70 $\pm$ 5 & 770 $\pm$ 15 \\
1.5 & 83 $\pm$ 7 & 745 $\pm$ 20 \\
2.0 & 90 $\pm$ 6 & 675 $\pm$ 20 \\
\hline
\end{tabular}
\end{center}
\end{table}

\begin{figure}[tbh]
\begin{center}
\mbox{\epsfig{file=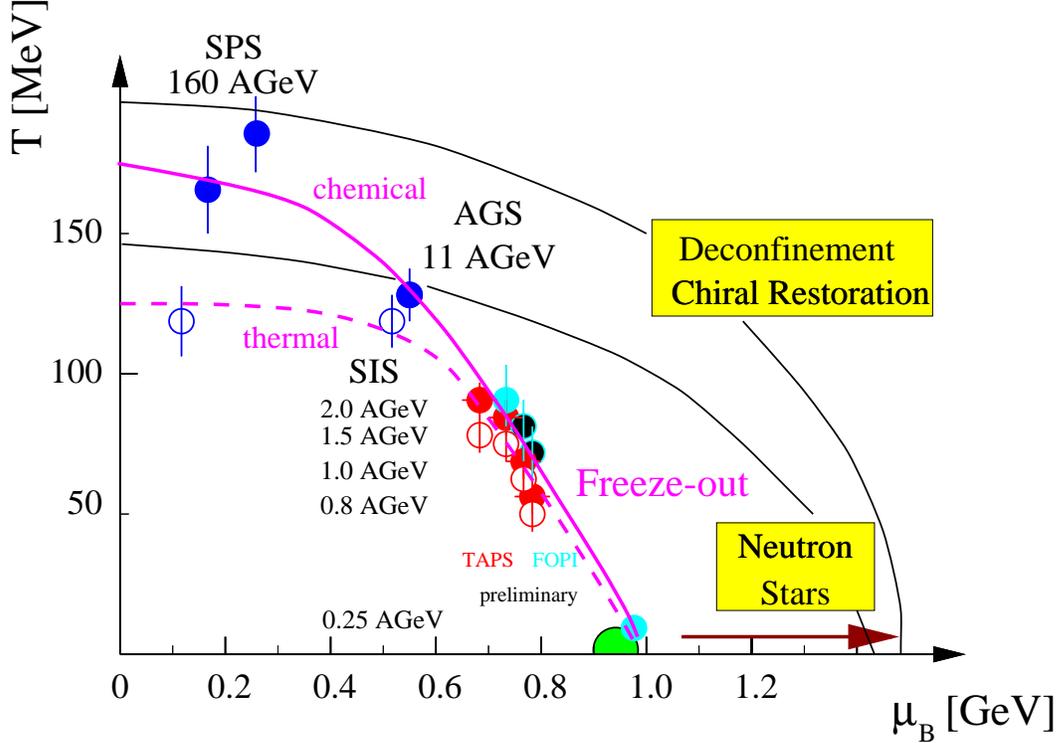,width=\textwidth}}
\end{center}
\caption[]{Phase diagram of hadronic matter: the temperatures $T$ and
baryon chemical potentials $\mu_B$ derived from yield ratios of particles
produced in nucleus--nucleus collisions at different incident energies.
Results of this work are plotted together with results taken from
\cite{BRA95,BRA96,HON96,STO98}. The solid curve through these data points
represents the curve for chemical freeze--out of hadronic matter.
Open data points represent parameter pairs for thermal freeze--out and are
connected by a corresponding freeze--out curve (dashed).}
\label{phase}
\end{figure}

Fig.~\ref{phase} shows our results within the phase diagram of
hadronic matter together with other $(T_C,\mu_B)$ values obtained from
particle--production experiments at SIS, AGS, and SPS energies
\cite{BRA95,BRA96,HON96,STO98}.
The data points are connected by a line to symbolize the chemical 
freeze--out curve of hadronic matter.
In contrast to the AGS and SPS results the deduced chemical freeze--out
parameters at SIS energies are far below the phase boundary between a 
hadron gas and a quark--gluon plasma.
Results dealing with thermal freeze--out, which are also shown in
Fig.~\ref{phase}, are discussed in the following chapter.

\begin{figure}[tbh]
\begin{center}
\mbox{\epsfig{file=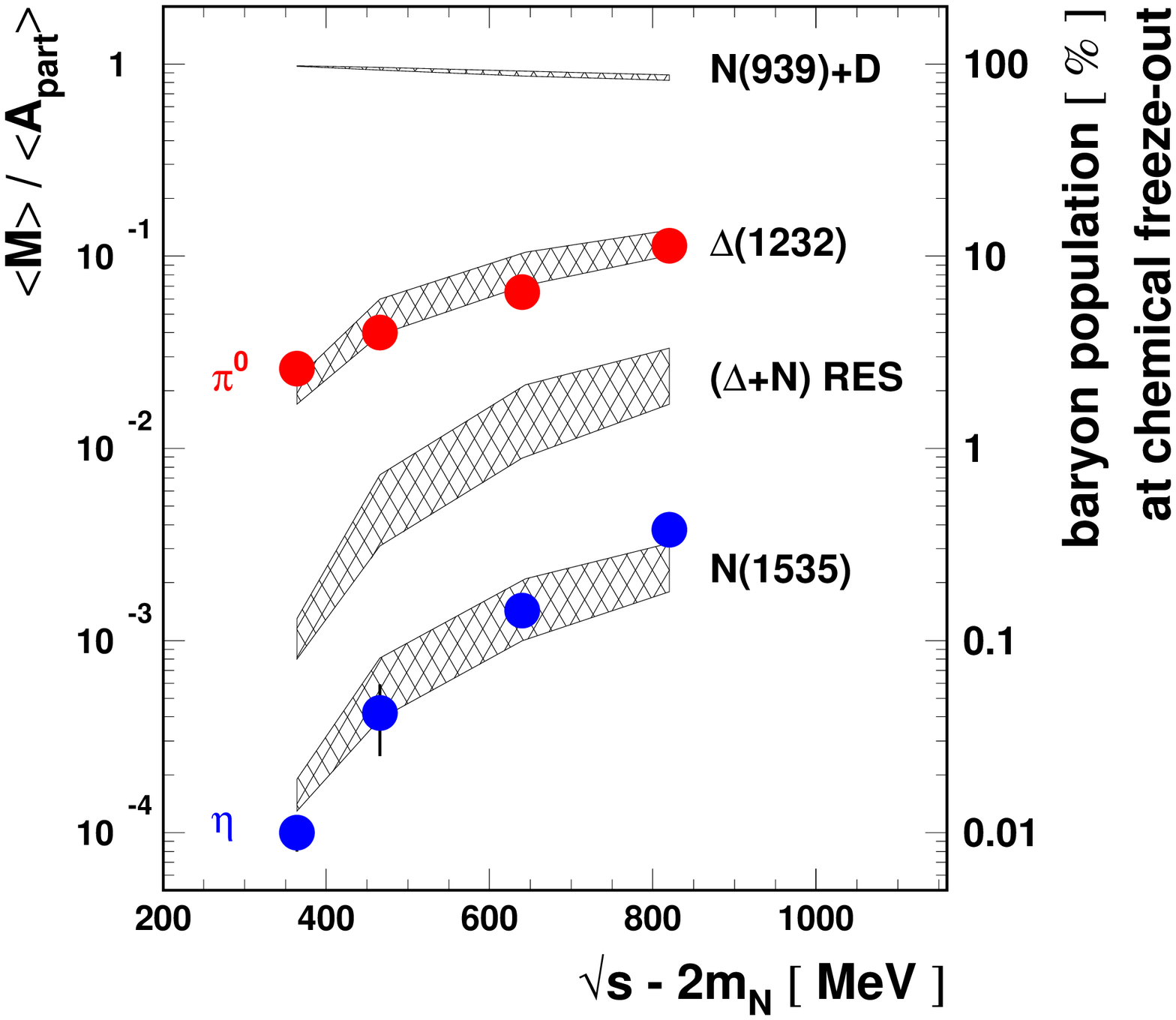,width=\textwidth}}
\end{center}
\caption[]{Inclusive $\pi^0$ and $\eta$--meson yields as function of the
energy available in the nucleon--nucleon system (left scale). The data points
represent system--size averaged values from various collision systems (see 
Tab.~\ref{data}). In addition, the hatched bands show the relative populations 
of nucleons and deuterons (summed), the $\Delta(1232)$ resonance, the 
$N(1535)$ resonance, and the sum of all remaining $\Delta$ and $N$ resonances
(right scale), as obtained from the hadron--gas model described in the text.
Only one third of the $\Delta(1232)$ resonances decays into neutral 
pions. Considering that due to the limited available energy the $N(1535)$ 
resonance is predominantly populated at masses below the nominal
mass of 1535~MeV it is comprehensible that less than 50~\%, which is
the branching ratio at the nominal mass, of these resonances decay
into the $\eta$ channel.}
\label{comp}
\end{figure}

Having determined the chemical freeze--out parameters one can study
the chemical composition of the hadronic fireball within the 
hadron--gas model.
Fig.~\ref{comp} shows the calculated relative populations of baryon
species at chemical freeze--out as a function of the energy available
in the nucleon--nucleon system together with the system--size
averaged relative meson multiplicities $<M_{\pi^0,\eta}>/<A_{part}>$
as given in Tab.~\ref{data}.
The hatched areas correspond to the relative populations of nucleons and
deuterons (summed), $\Delta(1232)$ and $N(1535)$ resonances as the main
sources of pions and $\eta$ mesons, respectively, and the sum of all 
remaining $\Delta$ and $N$ resonances. 
With increasing available energy the fraction of baryons excited into
resonance states at chemical freeze--out grows from $\approx$2~\%
at 0.8$A$~GeV beam energy to $\approx$15~\% at 2.0$A$~GeV.
The $\Delta(1232)$ resonance is populated most abundantly since it is
the lowest mass baryon resonance.
The ratio of heavier resonances to the $\Delta(1232)$ increases from
$\approx$5~\% to $\approx$15~\% if one goes from low to high energies.
Furthermore, it is obvious that the decay of resonances still excited 
at chemical freeze--out is not sufficient to account for the meson
multiplicities observed asymptotically in the detector.
Consequently, a sizeable fraction of resonance decays occurs before 
freeze--out is reached:
at chemical freeze--out $\approx$60~\% of the pions and $\approx$90~\%
of the $\eta$ mesons are not bound in baryon resonances but are already
present as free mesons.

Finally, one can extrapolate the resonance population at chemical
freeze--out to the population in the high--density phase of the collision.
Guided by microscopic model calculations \cite{TEI97,WEB96} which quite 
consistently give a maximum baryon density of
$\rho_{max}$~$\approx$~2.5~$\rho_0$ and which predict ratios around 0.6 for
the number of $\Delta(1232)$ resonances present at maximum compression
relative to the number of asymptotically observed pions, one obtains an
initial resonance density of $0.6 \times 0.113 \times 3 
\approx$~0.2~$\rho_{max}$~$\approx$~0.5~$\rho_0$ at 2$A$~GeV beam energy.
Since in the high--density phase of the collision the baryon--resonance
density becomes comparable to the nuclear groundstate density, hadronic
matter in that stage is often referred to as resonance matter 
\cite{EHE93,MOS93}.

\section{Particle Spectra and Thermal Freeze--Out in a Thermal Model}

While particle yields contain information about chemical freeze--out
parameters as discussed in the preceeding chapter, particle spectra 
reflect the nucleus--nucleus collision process at the stage of thermal 
freeze--out in our simple sudden freeze--out scenario.
Even if the system reaches global thermal equilibrium at freeze--out
the spectra, however, do not follow pure Boltzmann distribution 
functions, but they are modified by resonance decays and they furthermore
reflect collective flow  phenomena.

We present here an analysis of the midrapidity $\pi^0$ and $\eta$--meson
spectra within the blast model proposed by Siemens and Rasmussen
\cite{SIE79}.
In this model, a fireball in thermal equilibrium expands isotropically.
All particles in the fireball share a common temperature $T$ and 
radial--flow velocity $\beta_r$, which are the two parameters that
characterize the fireball at thermal freeze--out.
The transverse--mass spectra are then described by

\begin{equation}
\frac{1}{m_t^2} \frac{d^2N}{dm_t dy} \propto
\cosh y \exp(-\gamma_r E/T) 
\left[\left(\gamma_r+\frac{T}{E}\right)\frac{\sinh\alpha}{\alpha} -
\frac{T}{E}\cosh\alpha\right]
\label{blast}
\end{equation} 

where $\gamma_r = 1/\sqrt{1-\beta_r^2}$, $\alpha = \left(\gamma_r \beta_r
p\right)/T$, and $E$, $p$, and $y$ are the total energy, momentum and
rapidity of the considered particle in the center--of--mass system.  
The modification of the spectra compared to pure Boltzmann distributions
becomes more significant the heavier the considered particle species is.

\begin{figure}[tbh]
\begin{center}
\mbox{\epsfig{file=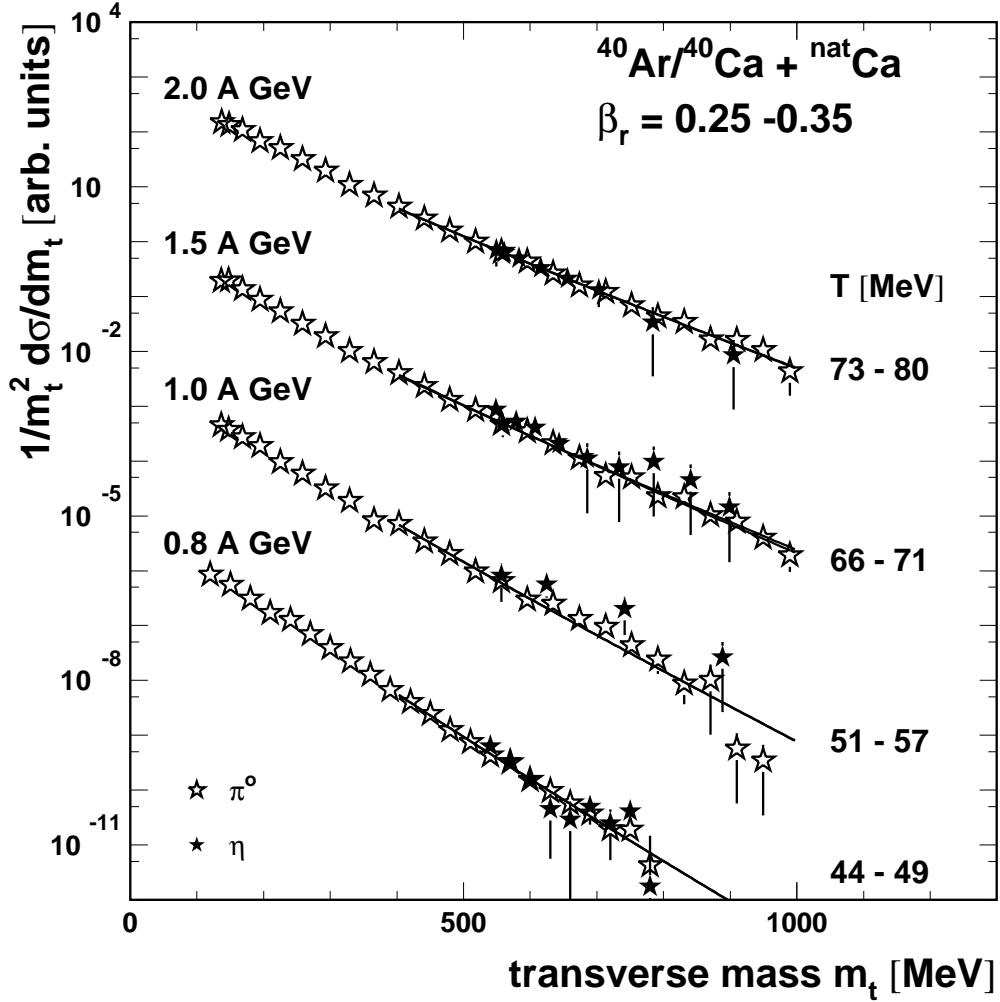,width=\textwidth}}
\end{center}
\caption[]{Transverse--mass spectra of $\pi^0$ and $\eta$ mesons
measured in the system $^{40}$Ar/$^{40}$Ca+$^{nat}$Ca 
\cite{BER94,SCH94,MAR97,VOG98} at beam energies of 0.8, 1.0, 1.5, and 
2.0$A$~GeV. The cross sections are scaled in arbitrary units. For each
beam energy two lines are drawn which represent fits to the $\pi^0$ data
above transverse masses of 400~MeV using Eq.~\ref{blast} with fixed 
transverse--expansion velocities of $\beta_r = 0.25$ and 0.35, respectively.
On the scale chosen a difference between the two fits is not visible.}
\label{arcaspec}
\end{figure}

Since pions and $\eta$ mesons have low masses their spectra are not very 
sensitive to the radial--flow velocity.
Therefore a fit to the transverse--mass spectra using $T$ and $\beta_r$
both as free parameters allows to determine the freeze--out parameters
only with very large uncertainties.
To overcome this difficulty, we take $\beta_r$ from the systematics of 
transverse--flow velocities for different collision systems and bombarding 
energies as published in \cite{HER96}.
Values for $\beta_r$ between 0.25 and 0.35 are found to be a reasonable 
choice in the considered range of incident beam energies.
To determine the thermal freeze--out parameter $T$ we then fit the
measured midrapidity transverse--mass spectra of $\pi^0$ and $\eta$ mesons
using Eq.~\ref{blast}, where we vary only $T$ and keep $\beta_r$ constant. 
In the case of the pion spectra, we furthermore restrict the fit to
$m_t \ge 400$~MeV since the low $m_t$ region is expected to be modified
by resonance decays.
As an example Fig.~\ref{arcaspec} shows the meson $m_t$ spectra measured 
in the systems $^{40}$Ar/$^{40}$Ca + $^{nat}$Ca together with two fits
for each considered beam energy.
The fits corresponds to values of $\beta_r = 0.25$ and $\beta_r = 0.35$,
respectively.
Obviously, the distributions are consistent with a thermal freeze--out
scenario as assumed in the blast model.
This is also the case for the other investigated projectile--target 
combinations.

\begin{figure}[tbh]
\begin{center}
\mbox{\epsfig{file=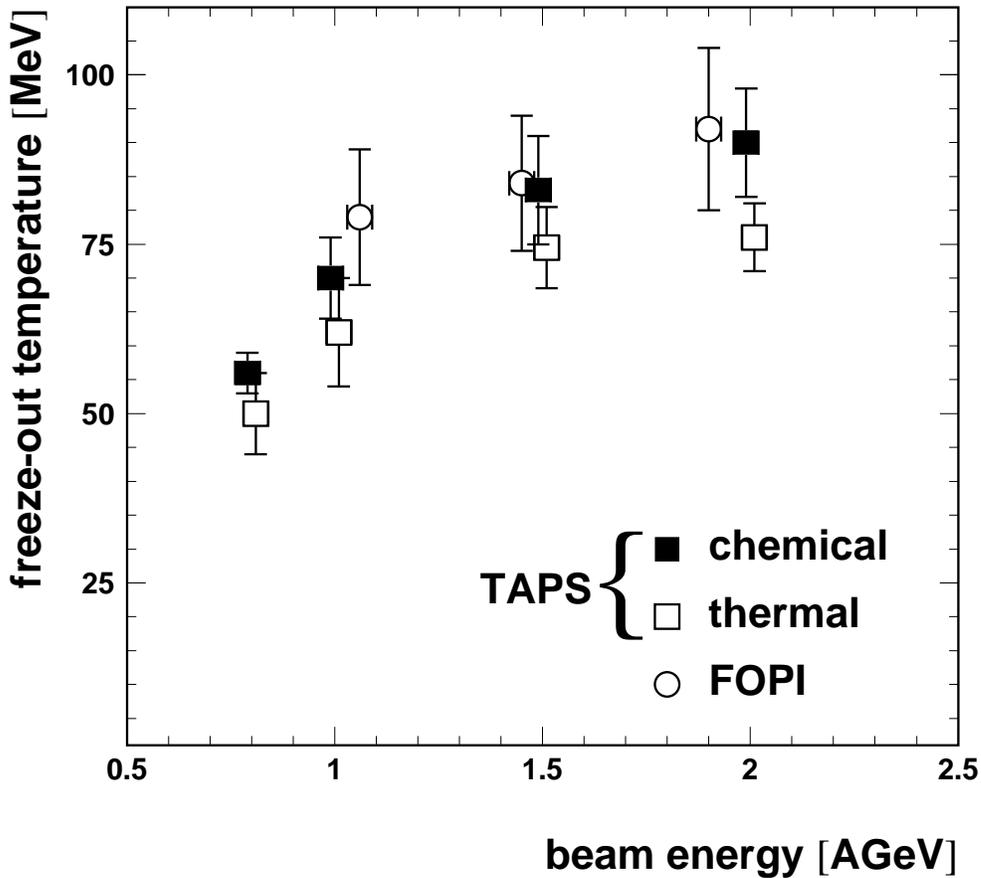,width=\textwidth}}
\end{center}
\caption[]{System--size averaged thermal and chemical freeze--out 
temperatures, respectively, as function of the beam energy. Thermal
freeze--out temperatures taken from \cite{HON98} are shown for 
comparision.}
\label{comptherm}
\end{figure}

Fig.~\ref{comptherm} shows the thermal freeze--out temperatures,
again averaged over system size, as a function of the incident energy,
together with the chemical freeze--out temperatures as obtained in
the previous chapter.
In addition, temperatures obtained in a blast--model analysis of
$\pi^-$, proton, and deuteron spectra from the FOPI collaboration
\cite{HON98} are shown for comparision.
With increasing beam energy the thermal as well as the chemical
freeze--out temperatures grow.
Within the error bars, chemical and thermal freeze--out temperatures
overlap and agree nicely with the FOPI results.
Our analysis, however, shows some trend towards lower temperatures at
thermal freeze--out which one could expect since thermal freeze--out
should not occur before chemical freeze--out.

Finally, we want to include our results on thermal freeze--out
parameters into the phase diagram of hadronic matter as shown in
Fig~\ref{phase}.
To this aim one has to estimate the baryon chemical potential $\mu_B$
at the moment of thermal freeze--out.
If one assumes an isentropic expansion of the collision zone one can 
make use of the fact that the ratio $T/\mu_B$ remains constant in
that case.
The baryon chemical potential at thermal freeze--out may then be calculated 
from the one at chemical freeze--out just by multiplication with
the ratio of the temperatures.
The resulting thermal freeze--out parameters are included in 
Fig.~\ref{phase} together with a curve representing the thermal 
freeze--out of hadronic matter.
While at SIS energies thermal and chemical freeze--out seem to almost
coincide, at higher energies the freeze--out parameters clearly indicate 
that thermal freeze--out reflects a later stage of the collision process
as lower temperatures compared to chemical freeze--out are deduced.

\section{Conclusion}

Neutral--pion and $\eta$-meson data measured near midrapidity in the
energy range from 0.8$A$~GeV to 2.0$A$~GeV are consistent with the
assumption that the hadronic fireball, which evolves during relativistic
nucleus--nucleus collisions, may be in hadrochemical or thermal equilibrium 
at the moment of chemical or thermal freeze--out, respectively.

In contrast to ultrarelativistic collisions studied at AGS and SPS,
chemical and thermal freeze--out seem to occur at almost the same
stage of the expansion process at SIS energies.
Another difference is that the freeze--out parameters at SIS are far 
below the phase boundary between a hadron gas and a quark--gluon
plasma.
This implies that the early stage of an ultrarelativistic nucleus--nucleus 
collision may come close to that phase transition while at SIS the early,
high--density phase is described best by the term resonance matter.

Neutral--meson yields and spectra reflect only one facet of the very 
complex process of a relativistic nucleus--nucleus collision.
With these observable alone it is not possible to decide whether 
chemical or thermal equilibrium, respectively, is actually reached 
during any stage of such a collision.
Further experiments and analyses investigating complementary probes 
are necessary to clarify to what extent nucleus--nucleus collisions
in the SIS--energy range can be described within the framework of
equilibrium models.

\section*{Acknowledgement}

It is a pleasure to gratefully acknowledge helpful discussions with
many colleagues, in particular, P.~Braun--Munzinger, J.~Cleymans,
P.~Danielewicz, R.~Holzmann, V.~Metag, P.~Senger and R.S.~Simon.


\begin{thebibliography}{99}
\bibitem{STO86} R.~Stock: Phys. Rep. 135 (1986) 259 and references therein
\bibitem{CAS90} W.~Cassing et al.: Phys. Rep. 188 (1990) 363
\bibitem{AIC91} J.~Aichelin: Phys. Rep. 202 (1991) 233
\bibitem{MAR94} T.~Maruyama et al.: Nucl. Phys. A 573 (1994) 653
\bibitem{BAS95} S.A.~Bass et al.: Phys. Rev. C 51 (1995) 3343
\bibitem{EHE93} W.~Ehehalt et al.: Phys. Rev. C 47 (1993) 2467
\bibitem{MET93} V.~Metag: Prog. Part. Nucl. Phys. 30 (1993) 75
\bibitem{MOS93} U.~Mosel, V.~Metag: Phys. Bl. 49 (1993) 426
\bibitem{KAH96} S.H.~Kahana et al.: Ann. Rev. Nucl. Part. Sci. 46 (1996) 31
\bibitem{ABR96} M.C.~Abreu et al.: Nucl. Phys. A 610 (1996) 404c
\bibitem{REI97} W.~Reisdorf et al.: Nucl. Phys. A 612 (1997) 493
\bibitem{HON98} B.~Hong et al.: Phys. Rev. C 57 (1998) 244
\bibitem{CLE93} J.~Cleymans and H.~Satz: Z. Phys. C 57 (1993) 135
\bibitem{BRA95} P.~Braun-Munzinger et al.: Phys. Lett. B 344 (1995) 43
\bibitem{BRA96} P.~Braun-Munzinger et al.: Phys. Lett. B 365 (1996) 1
\bibitem{SOL97} J.~Sollfrank: J. Phys. G 23 (1997) 1903
\bibitem{HON96} B.~Hong: in Proceedings of "Heavy Ion Physics at Low,
Intermediate and Relativistic Energies using 4$\pi$ Detectors", 
Poiana Brasov, Romania, (1996) 304
\bibitem{CLE97} J.~Cleymans et al.: nucl-th/9711066
\bibitem{PFE93} M.~Pfeiffer: PhD thesis, University Gie{\ss}en, 1993, 
unpublished
\bibitem{BER94} F.-D.~Berg et al.: Phys. Rev. Lett. 72 (1994) 977
\bibitem{SCH94} O.~Schwalb et al.: Phys. Lett. B 321 (1994) 20
\bibitem{MET94} V.~Metag: private communication
\bibitem{SCH95} A.~Schubert et al.: Nucl. Phys. A 583 (1995) 385 
                                    and private communication
\bibitem{APP97} M.~Appenheimer: PhD thesis, University Gie{\ss}en, 1997,
unpublished
\bibitem{AVE97} R.~Averbeck et al.: Z. Phys. A 359 (1997) 65
\bibitem{MAR97} A.~Mar\'{\i}n et al.: Phys. Lett. B 409 (1997) 77
\bibitem{MAR98} G.~Mart\'{\i}nez et al.: to be published
\bibitem{VOG98} P.~Vogt: PhD thesis, University Groningen, 1998, unpublished
\bibitem{WOL98} A.R.~Wolf et al.: submitted to Phys. Rev. Lett.
\bibitem{MUE97} C.~M\"untz et al.: Z. Phys. A 357 (1997) 
\bibitem{PEL97} D.~Pelte et al.: Z. Phys. A 357 (1997) 215
\bibitem{SIM86} L.~Simi\'{c}: Phys. Rev. D 34 (1986) 692
\bibitem{STA92} J.~Stachel, G.R.~Young: Ann. Rev. Nucl. Part. Sc. 42 (1992) 537
\bibitem{BOU76} M.~Bourquin, J.M~Gaillard: Nucl. Phys. B 114 (1976) 334
\bibitem{AGU91} M.~Aguilar-Benitez et al.: Z. Phys. C 50 (1991) 405
\bibitem{BAS93} S.A.~Bass et al.: Phys. Rev. Lett. 71 (1993) 1144
\bibitem{TEI97} S.~Teis et al.: Z. Phys. A 356 (1997) 421
\bibitem{BAL70} R.~Balian and C.~Bloch: Ann. Phys. 70 (1970) 401
\bibitem{JAQ84} H.R.~Jaqama et al.: Phys. Rev. C 29 (1984) 2067
\bibitem{BAR96} R.M.~Barnett et al.: Phys. Rev. D 54 (1996) 1
\bibitem{KOC84} J.H.~Koch et al.: Ann. Phys. 154 (1984) 99
\bibitem{STO98} R.~Stock: Nucl. Phys. A 630 (1998) 
\bibitem{WEB96} H.~Weber: Diploma thesis, University Frankfurt, 1996,
unpublished 
\bibitem{SIE79} P.J.~Siemens and J.O.~Rasmussen: Phys. Rev. Lett. 42 (1979) 880
\bibitem{HER96} N.~Herrmann: Nucl. Phys. A 610 (1996) 49c
\end{thebibliography}
\end{document}